%% file: hadron2011.tex
\documentclass[a4paper,11pt]{article}

\usepackage{contribution}

\input{econfmacros}
\input{contributionmacros}

\begin{document}

\input{contribution}

\end{document}

%% file: econfmacros.tex
\newcommand{\weblink}[2][]{%
    \ifthenelse{\equal{#1}{}}%
    {\textnormal{\url{#2}}}%
    {\textnormal{\href{#2}{#1}}}%
}

\newcommand{\acknowledgements}[1]{%
  \bigskip\bigskip
  \textsf{\textbf{\Large Acknowledgements}} \\[2ex]
  {#1}
  \bigskip
}

\def\be{\begin{equation}}
\def\eeq#1{\label{#1}\end{equation}}
\def\ee{\end{equation}}

\def\bea{\begin{eqnarray}}
\def\eeqa#1{\label{#1}\end{eqnarray}}
\def\eea{\end{eqnarray}}

\def\ba{\begin{array}}
\def\ea{\end{array}}

\def\bc{\begin{center}}
\def\ec{\end{center}}

\let\bar=\overbar

\def\Dslash{\not{\hbox{\kern-4pt $D$}}}
\def\dslash{\not{\hbox{\kern-2pt $\del$}}}

\def\ee{e^+e^-}

\def\msb{{\bar{\ssstyle M \kern -1pt S}}}

%% file: contributionmacros.tex
\newcommand{\contribution}[7][]{%
  \clearpage
  \thispagestyle{plain}
  \ifthenelse{\equal{#1}{}}
  {\hypersetup{pdftitle={#2}}}
  {\hypersetup{pdftitle={#1}}}
  \hypersetup{pdfauthor={{#3} {#4}}}
  {\centering\normalfont\LARGE\bfseries\sffamily #2 \par\nobreak}
  \lhead{}
  \chead{%
    \textit{\footnotesize XIV International Conference on Hadron Spectroscopy
      (\weblink[\textit{hadron2011}]{http://www.hadron2011.de}), 13-17 June 2011, Munich, Germany}%
  }
  \rhead{}
  \bigskip
  \begin{center}
    {#3} {#4}\ifthenelse{\equal{#6}{}}{}{\footnote{\weblink[#6]{mailto:#6}}}
    \ifthenelse{\equal{#7}{}}{}{#7} \\
    \textit{#5}
  \end{center}
  \bigskip
}

\renewcommand{\abstract}[1]{%
  \begin{center}
    \begin{minipage}{0.85\textwidth}
      \begin{footnotesize}
        #1
      \end{footnotesize}
    \end{minipage}
  \end{center}
  \bigskip
}

%% file: contribution.tex
%
%
%
%
%
{  



%

\contribution[Strangeness magnetic moments]  
{Strangeness magnetic moments of $N$ and $\Delta$}  
{Harleen}{Dahiya}   
{Department of Physics, Dr. B.R. Ambedkar National Institute of
Technology, Jalandhar-144011, India} {dahiyah@nitj.ac.in} {and
Neetika Sharma}

\abstract{%
We have calculated the strangeness contribution to the magnetic
moments of the nucleon and $\Delta$ decuplet baryons in the chiral
constituent quark model with configuration mixing ($\chi$CQM$_{{\rm
config}}$) which is known to provide a satisfactory explanation of
the proton spin crisis and related issues. Our results are
consistent with the recent experimental observations. }
%

\section{Introduction}

The recent measurements by several groups SAMPLE  at MIT-Bates
\cite{sample}, G0  at JLab \cite{g0}, A4 at MAMI \cite{a4} and by
HAPPEX  at JLab \cite{happex} regarding the contribution of
strangeness to the electromagnetic form factors of the nucleon have
triggered a great deal of interest in finding the strangeness
magnetic moment of the proton ($\mu(p)^s$). The SAMPLE experiment
has observed $\mu(p)^s$ to be $0.37 \pm 0.26 \pm 0.20$ \cite{sample}
whereas G0 \cite{g0}, A4 \cite{a4} and HAPPEX \cite{happex} have
observed the combination of electric and magnetic form factors. It
is widely recognized that a knowledge about the strangeness content
of the nucleon would undoubtedly provide vital clues to the
non-perturbative aspects of QCD.

Chiral constituent quark model ($\chi$CQM) \cite{manohar} can yield
an adequate description of the quark sea generation through the
chiral fluctuations and is also successful in giving a satisfactory
explanation of proton spin crisis \cite{eichten}. Recently, it has
been shown chiral constituent quark model with configuration mixing
($\chi$CQM$_{{\rm config}}$) when coupled with the quark sea
polarization and orbital angular momentum through the Cheng-Li
mechansim \cite{cheng} is able to give an excellent fit
\cite{hdorbit} to the octet and decuplet magnetic moments. It,
therefore, becomes desirable to carry out the calculations of the
strangeness contribution to the magnetic moments of nucleon in the
$\chi$CQM$_{{\rm config}}$ in the light of some recent observations
\cite{sample,g0,a4,happex,PDG}. For the sake of completeness, we would
also like to calculate the strangeness contribution to the magnetic
moments of decuplet baryons $\mu(\Delta^{++})^s$,
$\mu(\Delta^{+})^s$, $\mu(\Delta^{o})^s$ and $\mu(\Delta^{-})^s$
which have not been observed experimentally.

\section{Chiral Constituent Quark Model}
The basic process in the $\chi$CQM formalism is the emission of a
Goldstone boson (GB) by a constituent quark which further splits
into a $q \overline{q}$ pair \cite{cheng,johan,hd}, for example, $
q_{\pm} \rightarrow {\rm GB}^{0}
  + q^{'}_{\mp} \rightarrow  (q \bar q^{'})
  +q_{\mp}^{'}$, where $q\overline{q}^{'}+q^{'}$ constitute the quark sea
\cite{cheng} and the $\pm$ signs refer to the quark helicities. The
effective Lagrangian describing interaction between quarks and a
nonet of GBs, consisting of octet and a singlet, can be expressed as
${\cal L}= g_8 {\bf \bar q}\left(\Phi+\zeta\frac{\eta'}{\sqrt 3}I
\right) {\bf q}=g_8 {\bf \bar q}\left(\Phi'\right) {\bf q}$, where
$\zeta=g_1/g_8$, $g_1$ and $g_8$ are the coupling constants for the
singlet and octet GBs, respectively, $I$ is the $3\times 3$ identity
matrix. The GB field which includes the octet and the
singlet GBs is written as \begin{eqnarray}
 \Phi' = \left( \begin{array}{ccc} \frac{\pi^0}{\sqrt 2}
+\beta\frac{\eta}{\sqrt 6}+\zeta\frac{\eta^{'}}{\sqrt 3} & \pi^+
  & \alpha K^+   \\
\pi^- & -\frac{\pi^0}{\sqrt 2} +\beta \frac{\eta}{\sqrt 6}
+\zeta\frac{\eta^{'}}{\sqrt 3}  &  \alpha K^0  \\
 \alpha K^-  &  \alpha \bar{K}^0  &  -\beta \frac{2\eta}{\sqrt 6}
 +\zeta\frac{\eta^{'}}{\sqrt 3} \end{array} \right) {\rm and} ~~~~q =\left( \begin{array}{c} u \\ d \\ s \end{array}
\right)\,. \end{eqnarray}

SU(3) symmetry breaking is introduced by considering $M_s >
M_{u,d}$ as well as by considering the masses of GBs to be
nondegenerate
 $(M_{K,\eta} > M_{\pi}$ and $M_{\eta^{'}} > M_{K,\eta})$
\cite{cheng,johan}. The parameter $a(=|g_8|^2$) denotes the
probability of chiral fluctuation  $u(d) \rightarrow d(u) +
\pi^{+(-)}$, $\alpha^2 a$, $\beta^2 a$ and $\zeta^2 a$
 respectively denote the probabilities of fluctuations
$u(d) \rightarrow s + K^{-(0)}$,  $u(d,s) \rightarrow u(d,s) +
\eta$ and $u(d,s) \rightarrow u(d,s) + \eta^{'}$.

\section{Magnetic moment}
The magnetic moment of a given baryon in the $\chi$CQM can be
expressed as $\mu(B)_{{\rm total}} = \mu(B)_{{\rm val}} +
\mu(B)_{{\rm sea}}$, where $\mu(B)_{{\rm val}}$ represents the
contribution of the valence quarks and $\mu(B)_{{\rm sea}}$
corresponding to the quark sea. Further, $\mu(B)_{{\rm sea}}$ can be
written as $\mu(B)_{{\rm sea}} = \mu(B)_{{\rm spin}} + \mu(B)_{{\rm
orbit}}$, where the first term is the magnetic moment contribution
of the $q^{'}$ coming from the spin polarization and the second term
is due to the rotational motion of the two bodies, $q^{'}$ and GB
and referred to as the orbital angular momentum by Cheng and Li
\cite{cheng}.

The strangeness contribution to the magnetic moment of the proton
$\mu(p)^s$ receives contributions only from the quark sea and is
expressed as $\mu(p)^s = \mu(p)^s_{{\rm spin}} + \mu(p)^s_{{\rm
orbit}}$ where $\mu(p)^s_{{\rm spin}}=\sum_{q=u,d,s}\Delta
q(p)^s_{{\rm sea}}\mu_q$ and $\mu(p)^s_{{\rm orbit}} = \frac{4}{3}
[\mu (u_+ \rightarrow s_-)]- \frac{1}{3} [\mu (d_+ \rightarrow
s_-)]$. Here, $\mu_q= \frac{e_q}{2 M_q}$ ($q=u,d,s$) is the quark
magnetic moment, $e_q$ and $M_q$ are the electric charge and the
mass respectively for the quark $q$ and $ \mu(q_+ \rightarrow s_-)
=\frac{e_{s}}{2M_q}\langle
l_q\rangle+\frac{e_{q}-e_{s}}{2M_{GB}}\langle l_{GB}\rangle$. The
quantities $(l_q, l_{GB})$ and $(M_q, M_{GB})$ are the orbital
angular momenta and masses of quark and GB, respectively. The
strangeness contribution to the magnetic moments of the neutron
$n(ddu)$ as well as the decuplet baryons $\Delta^{++}(uuu)$,
$\Delta^{+}(uud)$, $\Delta^{o}(udd)$ and $\Delta^{-}(ddd)$ can be
calculated similarly.

\section{Results and Discussion}

In Table \ref{mag}, we have presented the spin and orbital
contributions pertaining to the strangeness magnetic moment of the
nucleon and $\Delta$ baryons. From the Table one finds that the
present result for the strangeness contribution to the magnetic
moment of proton looks to be in agreement with the most recent
results available for  $\mu(p)^s$. On closer examination of the
results, several interesting points emerge. The strangeness
contribution to the magnetic moment is coming from spin and orbital
angular momentum of the quark sea with opposite signs. These
contributions are fairly significant and they cancel in the right
direction to give the right magnitude to $\mu(p)^s$, For example,
the spin contribution in this case is $-0.09 \mu_N$ and the
contribution coming from the orbital angular momentum is $0.05
\mu_N$. These contributions cancel to give a small value for
$\mu(p)^s$ $-0.03 \mu_N$ which is consistent with the other observed
results. Interestingly, in the case of $\mu(n)^s$, the magnetic
moment is dominated by the orbital part as was observed in the case
of the total magnetic moments \cite{hdorbit} however, the total
strangeness magnetic moment is same as that of the proton.
Therefore, an experimental observation of this would not only
justify the Cheng-Li mechanism \cite{cheng} but would also suggest
that the chiral fluctuations is able to generate the appropriate
amount of strangeness in the nucleon. For the sake of completeness,
we have also presented the results of $\mu(\Delta^{++})^s$,
$\mu(\Delta^{+})^s$, $\mu(\Delta^{o})^s$, $\mu(\Delta^{-})^s$ and
here also we find that there is a substantial contribution from spin
and orbital angular momentum. In general, one can find that whenever
there is an excess of $d$ quarks the orbital part dominates, whereas
when we have an excess of $u$ quarks, the spin polarization
dominates.

In conclusion, $\chi$CQM$_{{\rm config}}$ is able to provide a
fairly good description of the strangeness contribution to the
magnetic moment $\mu(p)^s$ and our result is consistent with the
latest experimental measurements as well as with the other
calculations. The constituent quarks and the weakly interacting
Goldstone bosons constitute the appropriate degrees of freedom in
the nonperturbative regime of QCD and the quark sea generation
through the chiral fluctuation is the key in understanding the
strangeness content of the nucleon.

\begin{table}[tb]
  \begin{center}
    \begin{tabular}{lcccc}
Baryon & Data  &  $\mu(B)^s_{{\rm spin}}$ & $\mu(B)^s_{{\rm orbit}}$ & $\mu(B)^s$ \\   \hline
 $p$ &$ 0.37 \pm 0.26 \pm 0.20$ \cite{sample} & $-0.09$ & 0.06 & $-0.03$
\\

$n$& $-$ & 0.06 & $-0.09$ & $-0.03$\\

$\Delta^{++}$ & $-$  & $-0.29$ & 0.18 & $-0.11$ \\

$\Delta^{+}$ & $-$  & $-0.14$ &  0.11 & $-0.03$ \\

$\Delta^{o}$ & $-$  &   $-0.04$ & $-0.03$ &  $-0.07$ \\

$\Delta^{-}$ & $-$ &  $-0.09$ & 0.15 & $0.06$\\

    \end{tabular}
    \caption{The calculated values of the strangeness contribution to
the magnetic moment of nucleon and $\Delta$ decuplet baryons in the
$\chi$CQM$_{{\rm config}}$.}
    \label{mag}
  \end{center}
\end{table}
%

\acknowledgements{%
  H.D. would like to thank the organizers of Hadron2011 and DAE-BRNS, Government of
India, for financial support. }


%

}  
